\documentclass[twocolumn,showpacs,preprintnumbers,superscriptaddress,prc,nofootinbib,floatfix]{revtex4}
\usepackage[unicode]{hyperref}
\usepackage{amsmath}
\usepackage{amsfonts}
\usepackage{amssymb}
\usepackage{graphicx}
\usepackage{dcolumn}
\usepackage{bm}

\newcommand{\wigner}[6]{\left(\begin{array}{ccc} #1 & #2 & #3 \\ #4 & #5 & #6 \end{array}\right)} 

\begin{document}

\title{Spectral properties of a tractable collective Hamiltonian}
\author{S. De Baerdemacker}
\email{stijn.debaerdemacker@ugent.be}
\affiliation{Universiteit Gent, Vakgroep Subatomaire- en Stralingsfysica, Proeftuinstraat 86, B-9000 Gent, Belgium}
\affiliation{Department of Physics, University of Toronto, Toronto, Ontario M5S 1A7, Canada} %
\author{K. Heyde}%
\affiliation{Universiteit Gent, Vakgroep Subatomaire- en Stralingsfysica, Proeftuinstraat 86, B-9000 Gent, Belgium}
\author{V. Hellemans}
\affiliation{Universiteit Gent, Vakgroep Subatomaire- en Stralingsfysica, Proeftuinstraat 86, B-9000 Gent, Belgium}
\affiliation{Universit\'e Libre de Bruxelles, Service de Physique Nucl\'eaire Th\'eorique, B-1050 Bruxelles, Belgium}%
\date{\today}

\begin{abstract}
The spectral properties of a tractable collective model Hamiltonian are studied.  The potential energy is truncated up to quartic terms in the quadrupole deformation variables, incorporating vibrational, $\gamma$-independent rotational and axially deformed rotational structures.  These physically significant limits are analysed in detail and confronted with well-established approximation schemes.  Furthermore, transitional Hamiltonians in between the limits are presented and discussed.  All results are obtained within a recently presented Cartan-Weyl based framework to calculate $SU(1,1)\times SO(5)$ embedded quadrupole collective observables. 
\end{abstract}

\maketitle

\section{Introduction}
The low-lying nuclear structure properties of heavy and medium-heavy atomic nuclei away from the shell closures are dominated by collective modes of motion \cite{rowe:70,bohr:98}.  Although contemporary nuclear shell-model calculations have pushed the barriers of computational feasibility with the advent of large-scale shell-model schemes \cite{caurier:07}, this region of the nuclear chart is still unaccessed territory for conventional shell-model calculations.  This is mainly due to the large degree of collectivity, requiring an exceeding amount of shell-model configurations with respect to the convergence of the ground- and first excited states.  To describe the structure of these atomic nuclei, one can use (relativistic) mean-field based models \cite{vautherin:70,gogny:73,walecka:74,bender:03,ring:96} or truncated shell-model descriptions, such as e.g. the symmetry-based Interacting Boson Model (IBM)\cite{iachello:87}.  These models are remarkably successful in describing a diversity of collective structures, such as vibrational and rotational bands or beyond (triaxiality, shape coexistence, ...).

The variety of collective structures arising in the IBM triggered numerous studies on the geometrical properties of its Hamiltonian as a function of the model parameters.  For this purpose, the coherent state mean-field approximation of the many-body problem \cite{ginocchio:80,dieperink:80,bohr:80} is very well suited as it couples the fundamental $s$ and $d$ bosons of the IBM to a quadrupole deformation field $\alpha$.  The global minimum of the energy surface constructed from this coherent state gives a good approximation of the ground-state energy of the Hamiltonian and, in addition, we obtain a geometrical interpretation in terms of the quadrupole variables $\alpha$ at the minimum.

Eventually, the semi-classical limit approach of the coherent state formalism \cite{vanroosmalen:82} fully establishes the connection between the mean-field total-energy surfaces of the IBM and the potential-energy surfaces of the geometrical Bohr-Mottelson model (BMM)\cite{rainwater:50,bohr:52,bohr:53}.  As a major result, it has been demonstrated in a series of seminal papers \cite{iachello:00,iachello:01,iachello:03} that the critical points along the symmetry-transition lines of the IBM can approximately be described by introducing flat potentials in the BMM which are analytically solvable, the so-called $E(5)$, $X(5)$ and $Y(5)$ solutions.  This observation has instigated a number of recent theoretical as well as experimental studies on the topic of quantum shape phase transitions.  From a theoretical point of view, large effort has been put in the identification of analytically solvable potentials in the BMM \cite{fortunato:05a,bonatsos:07}, the investigation of the critical-point structure of the total-energy surfaces \cite{jolie:01,jolie:02,leviatan:03,iachello:04,leviatan:05,leviatan:06,cejnar:08} and the exploration of spectroscopic properties of the IBM along the transition paths \cite{cejnar:08,cejnar:03,rowe:04b,rosensteel:05,garciaramos:05,heinze:06,thiamova:06,fossion:07}.   From the experimental side, a vast number of studies have been devoted to the signatures of critical behaviour in atomic nuclei and we like to refer the reader to Ref.~\cite{casten:07} for a timely review.

Despite being based on different underlying physical principles, the IBM and BMM have a very similar algebraic structure \cite{rowe:05c}.  It is well-established that the 3 branching limits of the IBM can be associated with limiting structures in the BMM.  The $U(5)$ limit describes harmonic quadrupole vibrations around a spherical minimum, the $O(6)$ limit can be related to  $\gamma$-independent $\beta$-deformed rotational motion and the $SU(3)$ limit has a close connection with the rotation-vibration model (RVM) limit \cite{faessler:62,eisenberg:87} of the BMM.  

This paper aims to explore the spectral properties of a transparent though inclusive collective Hamiltonian in the BMM, capable of reproducing the 3 forementioned limits of collective structure.  A good candidate for this purpose is the tractable collective Hamiltonian \cite{caprio:03}, expressed as a polynomial up to $(\alpha\cdot\alpha)^2$ in the potential energy.  The use of the class of polynomial potentials is legitimate from a physics point of view since every realistic potential or interaction can be expressed as a Taylor expansion up to a sufficient degree within the region of particular interest.  The problem then reduces to the construction of a matrix representation of the Hamiltonian, which can be done making use of a harmonic oscillator algebra in any dimension \cite{rowe:05b} (see also Sect. \ref{section:collectivemodel}).  Moreover, the approach enables a clear-cut connection with the coherent state total-energy surfaces of the IBM.

The paper is organised as follows.  In Sect. \ref{section:collectivemodel}, we recapitulate the underlying concepts of the collective model and its implementation.  For this purpose, we will use a recently presented method which treats the collective model by means of a pure algebraic technique within the Cartan-Weyl basis \cite{debaerdemacker:07a,debaerdemacker:08}.  In Sect. \ref{section:phases} we will discuss the tractable collective Hamiltonian with its spectral properties for the physically relevant model parameter space and present our conclusions in Sect. \ref{section:conclusions}.
\section{The collective model}\label{section:collectivemodel}
The collective model is a macroscopic model in the sense that it considers the atomic nucleus as a macroscopic object with a well-defined surface, much alike a charged liquid drop.  It is assumed that the strong attractive nucleon-nucleon interaction roughly favours a spherical shape, so the radius $R(\theta,\phi)$ of the nucleus can be expressed as a multipole expansion around a spherical shape
\begin{equation}\label{collectivemodel:radius}
 R(\theta,\phi)=R_0\left(1+\sum_{l}\alpha_{l}\cdot Y^l(\theta,\phi)\right),
\end{equation}
with $R_0$ the mean radius, $Y^l_m(\theta,\phi)$ the spherical harmonics and $\alpha_{lm}$ the collective variables of multipole order $l$ and projection $m$.  The dot denotes invariant angular momentum coupling $a_l\cdot b_l=(-)^l\sqrt{2l+1}[a_lb_l]^{(0)}_0$.  Truncating the multipole expansion for the radius (\ref{collectivemodel:radius}) at quadrupole order, we obtain the equation of an ellipsoid, as long as the quadrupole variables $\alpha_{2\mu}$ are sufficiently small (we will abbreviate $\alpha_{2\mu}$ by $\alpha_{\mu}$ from this point).  The monopole variable $\alpha_{00}$ is fixed by imposing volume conservation and the dipole variables $\alpha_{1\mu}$ merely describe small shifts of the centre of mass.  From a quantum mechanical point of view, eq.~(\ref{collectivemodel:radius}) is to be interpreted as a dynamic rather than a static deformation of the atomic nucleus, the dynamics being determined by the Bohr Hamiltonian \cite{bohr:52,bohr:53}
\begin{equation}\label{collectivemodel:hamiltonian}
 \hat{H}_B=\hat{T}+V(\alpha),
\end{equation}
The quadrupole deformation around the spherical shape is assumed to be small, justifying the $SO(3)$-scalar Taylor expansion of the potential around $\alpha_\mu=0$
\begin{align}\label{collectivemodel:potentialenergy}
 V(\alpha)&=c_2(\alpha\cdot\alpha)+c_3([\alpha\alpha]^2\cdot\alpha)+c_4(\alpha\cdot\alpha)^2\notag\\
    &\qquad +c_5([\alpha\alpha]^2\cdot\alpha)(\alpha\cdot\alpha) +c_6(\alpha\cdot\alpha)^3+\dots,
\end{align}
as has been proposed in the General Collective Model (GCM) by the Frankfurt group \cite{gneuss:71,hess:80a,hess:80b,troltenier:91,eisenberg:87}, which can be regarded as an extension of the BMM.  This extension is most manifest in the description of the kinetical energy
\begin{equation}\label{collectivemodel:kineticenergy}
 \hat{T}=\tfrac{1}{2B_2}(\hat{\pi}\cdot\hat{\pi})+B_3([\hat{\pi}\alpha]^2\cdot\hat{\pi}+\textrm{h.c.})+\dots,
\end{equation}
with the canonic momenta $\hat{\pi}$ defined by means of the standard Heisenberg-Weyl commutation relations (we will omit the operator sign not to overload the notation)
\begin{equation}
 [\pi_\mu,\alpha_\nu]=-i\hbar\delta_{\mu\nu},\quad[\pi_\mu,\pi_\nu]=0,\quad [\alpha_\mu,\alpha_\nu]=0.
\end{equation}
In the present work, we concentrate on a truncated version of the collective Hamiltonian, as has been discussed by Caprio \cite{caprio:03}.  Here, the potential energy (\ref{collectivemodel:potentialenergy}) consists of terms up to $(\alpha\cdot\alpha)^2$, which is sufficient to describe vibrational, $\gamma$-independent and axially deformed rotational structures.  The kinetic energy term (\ref{collectivemodel:kineticenergy}) is chosen in line with the standard quadratic expression, as proposed by Bohr and Mottelson \cite{bohr:52,bohr:53}.

In general, the Hamiltonian (\ref{collectivemodel:hamiltonian}) is not analytically solvable, so a suitable basis is required for numerical diagonalisation.  The five dimensional (5D) harmonic oscillator basis functions provide a genuine set of basis functions for this purpose as they can be generated by means of an $su(1,1)\times so(5)$ algebra, which is tailor-made for polynomial Hamiltonian problems in an $N$ dimensional Euclidean space \cite{rowe:05b}.  However, due to the dimensionality of the problem, the construction of an orthonormal set of 5D harmonic oscillator wavefunctions is a non-trivial and even ambiguous task if the quantum number $L$ of the angular momentum algebra $so(3)$ is to be preserved.  Therefore, a number of different techniques have been proposed and extensively discussed in the literature.  The basis functions up to $L=6$ have been determined first by B\`es by means of a coupled differential equation technique \cite{bes:58}.  Later on, the importance of the underlying algebraic structure of the 5D harmonic oscillator has been fully appreciated, leading towards the construction of explicit basis wavefunctions from basic tensorial building blocks \cite{corrigan:76,chacon:76,chacon:77} or from a coherent state formalism \cite{gheorghe:78}, with good orthonormality properties \cite{szpikowski:80,gozdz:80}.  Also the vector coherent state formalism \cite{rowe:94a,rowe:94b,rowe:95,turner:06} and, more recently, the algebraic tractable model \cite{rowe:04a,rowe:05a,turner:05b} are suited for the construction of quadrupole harmonic oscillator representations with good angular momentum.  Alternatively, we have shown recently \cite{debaerdemacker:07a,debaerdemacker:08} that it is possible to calculate the necessary matrix elements of all the physical observables with a pure algebraic technique {\it i.e.} without the need for explicit basis wavefunctions.  This technique transforms the angular momentum coupled $su(1,1)\times so(5)$ algebra (see eqs. \ref{collectivemodel:su11generators}-\ref{collectivemodel:angularmomentumcoupledO5generators}) into the Cartan-Weyl basis \cite{cartan:1894,wybourne:74,iachello:06} to calculate the matrix elements by means of an intermediate-state method.  The $su(1,1$) generators are given by the standard expressions \cite{ui:68,arima:76}
\begin{equation}
 B^\dag= \tfrac{1}{2}b^\dag\cdot b^\dag,\quad B=\tfrac{1}{2}\tilde{b}\cdot\tilde{b},\quad B_0=\tfrac{1}{4}(b^\dag\cdot\tilde{b}+\tilde{b}\cdot b^\dag),\label{collectivemodel:su11generators}
\end{equation}
with $b^\dag_\mu$ and $\tilde{b}_\mu$ the bosonic creation and annihilation operators
\begin{equation}
 b_\mu^\dag=\tfrac{1}{\sqrt{2}}(\sqrt{k}\alpha_\mu+\tfrac{i}{\sqrt{k}\hbar}\pi_\mu^\ast),\tilde{b}_\mu=\tfrac{1}{\sqrt{2}}(\sqrt{k}\alpha_\mu-\tfrac{i}{\sqrt{k}\hbar}\pi_\mu^\ast),
\end{equation}
and $k$ the spring constant of the 5D harmonic oscillator.  The Cartan-Weyl realisation of the $so(5)$ algebra is given by \cite{debaerdemacker:07a,corrigan:76}
\begin{align}
&X_+=-\tfrac{1}{5}(\sqrt{2}L_{+1}+\sqrt{3}O_{+1}),\quad  Y_+=-\tfrac{1}{\sqrt{5}}O_{+3},\notag\\
&X_-=\tfrac{1}{5}(\sqrt{2}L_{-1}+\sqrt{3}O_{-1}),\quad  Y_-=\tfrac{1}{\sqrt{5}}O_{-3},\notag\\
&X_0=\tfrac{1}{10}(L_0+3O_0),\quad Y_0=\tfrac{1}{10}(3L_0-O_0),\\
&T_{\frac{1}{2}\frac{1}{2}}=\tfrac{1}{\sqrt{10}}O_{+2},\quad  T_{-\frac{1}{2}\frac{1}{2}}=-\tfrac{1}{\sqrt{50}}(\sqrt{3}L_{+1}-\sqrt{2}O_{+1}),\notag\\
&T_{-\frac{1}{2}-\frac{1}{2}}=-\tfrac{1}{\sqrt{10}}O_{-2},\quad   T_{\frac{1}{2}-\frac{1}{2}}=\tfrac{1}{\sqrt{50}}(\sqrt{3}L_{-1}-\sqrt{2}O_{-1}),\notag
\end{align}
with the $L_\mu$ and $O_\mu$ respectively the angular momentum and octupole generators
\begin{align} 
&L_\mu=\sqrt{10}[b^\dag\tilde{b}]^{(1)}_\mu\equiv\tfrac{-i\sqrt{10}}{\hbar}[\alpha\pi^\ast]^{(1)}_\mu,\notag\\
&O_\mu=\sqrt{10}[b^\dag\tilde{b}]^{(3)}_\mu\equiv\tfrac{-i\sqrt{10}}{\hbar}[\alpha\pi^\ast]^{(3)}_\mu.\label{collectivemodel:angularmomentumcoupledO5generators}
\end{align}
From this definition it can be seen that the sets of generators $\{X_0,X_\pm\}$ and $\{Y_0,Y_\pm\}$ span the $so(4)\cong su(2)_X\times su(2)_Y$ algebra, which leads to the following group reduction 
\begin{equation}
 \underbrace{SU(1,1)}_{(n,v)}\times \underbrace{SO(5)}_{v}\supset \underbrace{SO(4)}_X\cong \underbrace{SU(2)_X}_{(X,M_X)}\times \underbrace{SU(2)_Y}_{(X,M_Y)}.
\end{equation}
The Cartan-Weyl basis $|nvXM_XM_Y\rangle$ of the $su(1,1)\times so(5)$ algebra is then defined by \cite{debaerdemacker:08}
\begin{align}
 &B_0|nvXM_XM_Y\rangle=\tfrac{1}{2}(2n+v+\tfrac{5}{2})|nvXM_XM_Y\rangle,\notag\\
 &\mathcal{C}_2[so(5)]|nvXM_XM_Y\rangle=v(v+3)|nvXM_XM_Y\rangle,\notag\\
 &\mathcal{C}_2[su(2)_{X,Y}]|nvXM_XM_Y\rangle=X(X+1)|nvXM_XM_Y\rangle,\notag\\
 &X_0|nvXM_XM_Y\rangle=M_X|nvXM_XM_Y\rangle\notag,\\
 &Y_0|nvXM_XM_Y\rangle=M_Y|nvXM_XM_Y\rangle,
\end{align}
with $(n,v)\in \mathbb{N}^2$ and $X=0,\frac{1}{2},1,\dots,\frac{v}{2}$.  $\{M_X,M_Y\}$ follow the standard $SU(2)$ reduction rules with respect to $X$.  It is convenient to calculate the matrix elements of $\alpha_\mu$ and $\pi_\mu$ (or equivalent $b^\dag_\mu$ and $\tilde{b}_\mu$) in this basis, as both operators carry good bitensorial properties within the $SU(2)_X\times SU(2)_Y$ reduction according to Racah's definition \cite{racah:42}. From this point, we will proceed with $\alpha_\mu$, though the results are generally valid for $\pi_\mu^\ast$, $b^\dag_\mu$ and $\tilde{b}_\mu$.
\begin{align}
&[X_{0},\alpha^{\lambda\lambda}_{\mu\nu}]=\mu \alpha^{\lambda\lambda}_{\mu\nu},\\
&[X_{\pm},\alpha^{\lambda\lambda}_{\mu\nu}]=\sqrt{(\lambda\mp\mu)(\lambda\pm\mu+1)}\alpha^{\lambda\lambda}_{\mu\pm1\nu},\\
&[Y_{0},\alpha^{\lambda\lambda}_{\mu\nu}]=\nu \alpha^{\lambda\lambda}_{\mu\nu},\\
&[Y_{\pm},\alpha^{\lambda\lambda}_{\mu\nu}]=\sqrt{(\lambda\mp\nu)(\lambda\pm\nu+1)}\alpha^{\lambda\lambda}_{\mu\nu\pm1},
\end{align}
where the collective variables $\alpha_\mu$ have been relabelled as
\begin{align}
 &\left\{\alpha^{00}_{00}=\alpha_0\right\}\\
 &\left\{\alpha^{\frac{1}{2}\frac{1}{2}}_{\frac{1}{2}\frac{1}{2}}=\alpha_2,\alpha^{\frac{1}{2}\frac{1}{2}}_{-\frac{1}{2}\frac{1}{2}}=\alpha_1,\alpha^{\frac{1}{2}\frac{1}{2}}_{\frac{1}{2}-\frac{1}{2}}=\alpha_{-1},\alpha^{\frac{1}{2}\frac{1}{2}}_{-\frac{1}{2}-\frac{1}{2}}=\alpha_{-2}\right\},\notag
\end{align}
which points out that the 5 projections of $\alpha_\mu$ can be subdivided into the 4 components of a $\{\frac{1}{2}\frac{1}{2}\}$ bispinor and a single $\{00\}$ biscalar.  These bitensorial properties facilitate the calculation of the matrix elements of the collective variables (and canonic conjugate momenta) considerably since we can revert to double-reduced matrix elements making use twice of the Wigner-Eckart theorem
\begin{align}\label{collectivemodel:doublereducedmatrixelements}
\langle n& v X M_XM_Y|\alpha^{\lambda\lambda}_{\mu\nu}|n^\prime v^\prime X^\prime M_X^\prime M_Y^\prime\rangle\notag\\
&=(-)^{\phi}\wigner{X}{\lambda}{X^\prime}{-M_X}{\mu}{M_X^\prime}\wigner{X}{\lambda}{X^\prime}{-M_Y}{\nu}{M_Y^\prime}\notag\\
&\qquad\times\langle nvX|||\alpha^\lambda|||n^\prime v^\prime X^\prime\rangle,
\end{align}
with $\phi=2X-M_X-M_Y$ and $\langle nvX|||\alpha^\lambda|||n^\prime v^\prime X^\prime\rangle$ the double-reduced matrix element.  Explicit analytic expressions for these double-reduced matrix elements can be obtained with an intermediate-state method and for further details of this derivations, we refer the reader to \cite{debaerdemacker:07a,debaerdemacker:08}.

The main difference between previously proposed methods and the present Cartan-Weyl based method is the embedding of the angular momentum algebra $so(3)$.  Whereas in the previous methods, explicit wavefunctions carrying good angular momentum are constructed to calculate the matrix elements, this step has been withdrawn in the Cartan-Weyl scheme.  So, regardless the unphysical nature of the quantum numbers $X$,$M_X$ and $M_Y$, the weight basis is unambiguously defined and leads towards an algebraically and computationally enhanced calculation of the matrix elements (see eq.~(\ref{collectivemodel:doublereducedmatrixelements})).  Once the matrix elements are computed, the Cartan-Weyl basis needs to be transformed back to the angular momentum basis, which can be done by diagonalising the operator
\begin{align}
 &L\cdot L=4X^2 -3[(X_0-3Y_0+\tfrac{1}{2})(X_0+Y_0+\tfrac{1}{2})-\tfrac{1}{4}]\notag\\
&\quad +4\sqrt{3}[T_{-\frac{1}{2}\frac{1}{2}}X_{-}+T_{\frac{1}{2}-\frac{1}{2}}X_{+}]+12T_{\frac{1}{2}-\frac{1}{2}}T_{-\frac{1}{2}\frac{1}{2}}.
\end{align}
After the rotation to the physical basis, the diagonalisation of the Hamiltonian proceeds within the separate angular momentum $L$ subspaces, similar to previously proposed methods.

The non-compactness of the $SU(1,1)$ underlying symmetry of the BMM requires extra care with respect to the diagonalisation of the Hamiltonian.  In principle, the Hamiltonian lives within an infinite dimensional Hilbert space, so we need to ascertain that the eigenvalues in the restricted space are sufficiently close to the exact values in the infinite space.  Therefore, we gradually enlarged the Hilbert space with subsequent harmonic oscillator shells until convergency is reached.  To find an optimised value for the basis harmonic oscillator spring constant, we applied the method of Margetan and Williams \cite{margetan:82}.  For all the calculations in this work, this resulted in a convergency of $1$ eV for all depicted states within 100 harmonic oscillator shells or less.
\section{The model space of the collective model}\label{section:phases}
We intend to study the spectral properties of the collective Hamiltonian
\begin{align}\label{modelspace:generalhamiltonian}
 \hat{H}&=\tfrac{1}{2B_2}\pi\cdot\pi+c_2(\alpha\cdot\alpha)\notag\\
&\quad+c_3([\alpha\alpha]^2\cdot\alpha)+c_4(\alpha\cdot\alpha)^2.
\end{align}
Expressed in the intrinsic framework, the potential energy $V(\alpha)$ is written as
\begin{equation}\label{modelspace:generalpotential}
 V(\beta,\gamma)=c_2\beta^2-\sqrt{\tfrac{2}{7}}c_3\beta^3\cos3\gamma+c_4\beta^4,
\end{equation}
which clearly points out that this potential can cover vibrational, $\gamma$-independent and axial rotational structures for different values of the parameters $\{c_2,c_3,c_4\}$.  We will refer to these potentials respectively as the $V_\textrm{spher}$, $V_{\gamma\textrm{-ind}}$ and $V_\textrm{rot}$ limiting potentials of the collective model. However, it should be emphasised that these limits are not to be considered as genuine branching limits associated with a symmetry algebra such as is the case \textit{e.g.}~in the IBM.  The algebra $su(1,1)\times so(5)$ in this case is an underlying, rather than a spectrum generating algebra, except in the case of the $SO(5)$ invariant $\gamma$-independent potentials of the collective model (eq.~(\ref{modelspace:generalpotential}) with $c_3=0$).  The specific choices for the parameters of the limiting potentials are presented in Table \ref{table:parameters}.
\begin{table}
 \begin{tabular}{lrrr}
 \hline
  & $c_2$ & $c_3$ & $c_4$\\
 \hline
 \hline
  $V_\textrm{spher}$        & \phantom{-}200 MeV & 0 MeV   & 0 MeV\\
  $V_{\gamma\textrm{-ind}}$ & -200 MeV           & 0 MeV   & 2500 MeV\\
  $V_\textrm{rot}$          & -200 MeV           & 700 MeV & 2500 MeV\\
  \hline
 \end{tabular}
 \caption{Parameters used in the potential (\ref{modelspace:generalpotential}) for the 3 limiting cases $V_\textrm{spher}$, $V_{\gamma\textrm{-ind}}$ and $V_\textrm{rot}$.}\label{table:parameters}
\end{table}
The value of $\hbar^2/B_2$ is chosen as 4 keV for all calculations, which agrees with values already used in realistic calculations \cite{petkov:95}.  
The motivation for these parameter-choices is twofold.  Within the restrictions of the given limit under study, the parameters have been chosen equal wherever possible.  This allows us to relate the differences arising in the structure to the particular parameter which has been varied.  On the other hand, although the sets of parameters give rise to schematic potentials, the connection with experimental observables was never neglected.  As a consequence, the three Hamiltonians can each act as a starting point for a profound study of atomic nuclei where typical fingerprints of the given limiting cases have been observed.
Though the parameters $\{c_2,c_3,c_4\}$ might appear to be rather large at first sight, one needs to realise that the collective coordinate $\beta$ describes small deformations.  Therefore, the parameters need to be significantly large to contribute to the general structure of the potential.
\subsection{The three limiting cases}
\subsubsection{Harmonic quadrupole oscillator}
\begin{figure}[!htb]
 \includegraphics[width=\columnwidth]{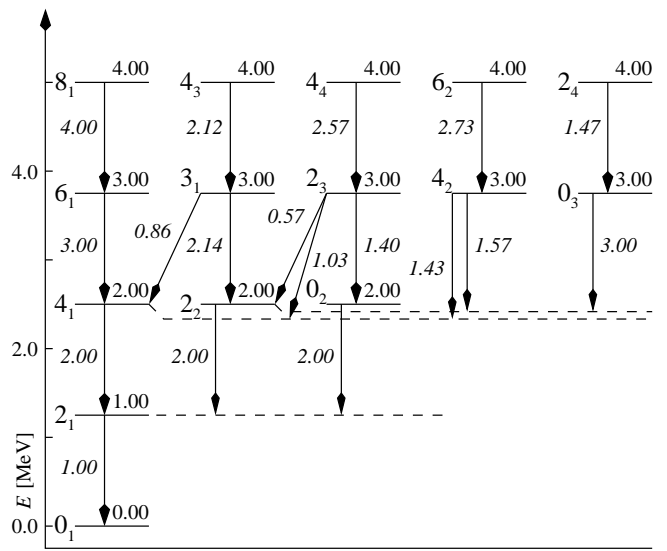}
 \caption{The level scheme and $B(E2)$ values of a harmonic oscillator potential (\ref{limits:vibhamiltonian}). Energy eigenvalues are given relative to the first $L=2$ state and $B(E2)$ values relative to $B(E2;2_1\rightarrow 0_1$).}\label{figure:vibspectrum}
\end{figure}

The harmonic oscillator is conceptually and computationally the simplest limit.  The Hamiltonian 
\begin{equation}\label{limits:vibhamiltonian}
 \hat{H}_\textrm{spher}=\tfrac{1}{2B_2}\pi\cdot\pi+V_\textrm{spher},
\end{equation}
reduces to the generator $B_0$ of the $su(1,1)$ algebra, which results in the well-known linear spectrum,
depicted in Fig.~\ref{figure:vibspectrum}.  Apart from the spectrum, all $B(E2;L_i\rightarrow L_f)$ transition values (relative to $B(E2;2_1\rightarrow 0_1)$) are presented up to harmonic oscillator shell $n=3$.  For $n=4$, only those states that can be organised into a band are shown (where a band is defined by following the cascade of large relative $B(E2)$ values).  For this reason, only the largest $B(E2)$ value is given for this set of $n=4$ states.  The excitation energy of the $2_1$ state is 1264.91 keV and the $B(E2;2_1\rightarrow 0_1)=0.00158$ in units relative to $(3ZeR_0^2)^2/(4\pi)^2$.  As all mentioned states throughout this paper have positive parity, the parity sign will not be explicitly denoted.
The energy spectrum can be determined solely using the $su(1,1)$ part of the algebra, whereas the calculation of the $B(E2)$ values also requires the $so(5)$ part.  This can be seen from the definition of the $B(E2)$ reduced transition probability
\begin{equation}\label{limits:be2}
 B(E2;L_i\rightarrow L_f)=e^2\sum_{\mu,M_f}|\langle L_fM_f|\hat{T}(E2)_\mu|L_iM_i\rangle|^2,
\end{equation}
with
\begin{equation}\label{limits:te2}
 \hat{T}(E2)_\mu=\tfrac{3ZR_0^2}{4\pi}\alpha_\mu^\ast.
\end{equation}
as $\alpha_\mu$ is a $v=1$ $SO(5)$-tensor connecting different $SO(5)$ representations with $\Delta v=1$.  $Z$ is the proton number and $e$ denotes the elementary charge of the proton.

Therefore, as the 5D harmonic oscillator has already been discussed numerously in the literature, it provides a reliable test for the matrix elements in this Cartan-Weyl reduction scheme.  As a result, we can proceed with confidence studying structures which can no longer be expressed as a genuine symmetry limit and for which a numerical treatment is required.
\subsubsection{\texorpdfstring{$\gamma$}--independent rotor}
The second limit to be considered is the $\gamma$-independent rotor, described by the potential $V_{\gamma\textrm{-ind}}$
\begin{equation}\label{limits:gammarothamiltonian}
 \hat{H}_{\gamma-\textrm{ind}}=\tfrac{1}{2B_2}\pi\cdot\pi+V_{\gamma-\textrm{ind}}.
\end{equation}
The occurrence of a global minimum at non-zero deformation $\beta_0=0$ makes this potential essentially different from the harmonic oscillator potential.  Consequently, this  potential is able to generate solutions which can be associated with definite deformations.

\begin{figure}[!htb]
 \includegraphics[width=\columnwidth]{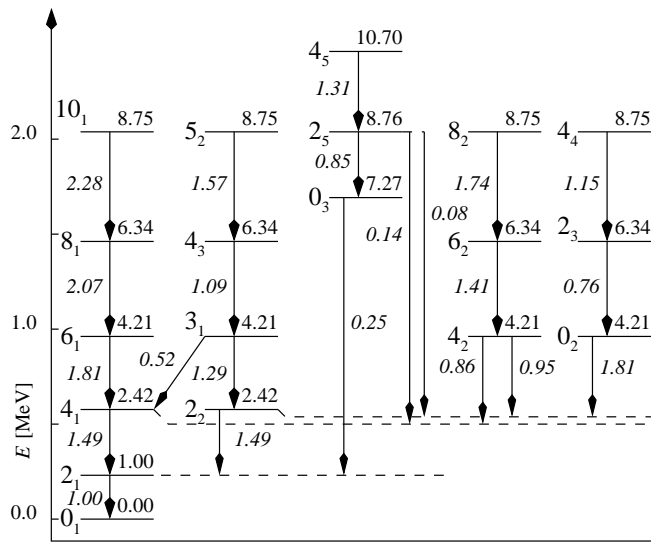}
 \caption{The level scheme and $B(E2)$ values of a $\gamma$-independent rotor potential (\ref{limits:gammarothamiltonian}).  Energy eigenvalues are given relative to the first $L=2$ state and $B(E2)$ values relative to $B(E2;2_1\rightarrow 0_1$).}\label{figure:gammaindspectrum}
\end{figure}

Although the Hamiltonian does not correspond to an algebraic solvable limit, still it exhibits some remarkable symmetry properties.  Similar to the case of the harmonic oscillator, the Hamiltonian is an $SO(5)$ invariant, leading towards degeneracies within a given representation $v$.  This is illustrated in Fig.~\ref{figure:gammaindspectrum} where for all bands except the one in the middle, the states with equal seniority $v$ have the same energy eigenvalue.  The main difference between $H_{\gamma\textrm{-ind}}$ and $H_\textrm{spher}$ with respect to the spectrum generating algebra $su(1,1)$ is that, although the Hamiltonian can entirely be written as a function of the generators of this group, it does no longer constitute a dynamical symmetry for the Hamiltonian.  In this particular case, it causes the $n\neq0$ states of the harmonic oscillator to be lifted in the spectrum.  This is illustrated in Fig.~\ref{figure:gammaindspectrum} by the band in the middle, which lies much higher in the excitation spectrum than the corresponding band of the harmonic oscillator (Fig.~\ref{figure:vibspectrum}).

Similar to Fig.~\ref{figure:vibspectrum}, all eigenstates with an excitation energy lower than or equal to the excitation energy of the $6_1$ state are depicted in Fig.~\ref{figure:gammaindspectrum}, together with all non-vanishing $B(E2)$ values.  For the higher-lying excited states, only those states are given that fit into a given band, as well as the corresponding intraband $B(E2)$ values to justify this classification.  All eigenvalues are again given relative to the $2_1$ state, which has an absolute excitation energy of 233.8~keV, and the $B(E2;2_1\rightarrow 0_1$)=0.00759, measured in units $(3ZeR_0^2)^2/(4\pi)^2$.

To validate the numerical calculations and to interprete the results, it is instructive to compare them with analytically solvable approximation schemes.  Such a scheme is provided by the displaced harmonic oscillator of the Wilets \& Jean (WJ) class of $\beta-\gamma$ decoupled potentials \cite{wilets:56}.  We can perform a Taylor expansion of the potential $V_{\gamma-\textrm{ind}}$ around the minimum $\beta_0=0.2$ of the potential, leading to a WJ displaced harmonic oscillator as long as the expansion is truncated up to second order.  The eigenvalues of this displaced harmonic oscillator are given by
\begin{equation}
 E_{n_\beta v}=\hbar\Omega[\omega^\prime(n_\beta+\tfrac{1}{2})+\tfrac{k}{2}(\beta_0^\prime(v)-\beta_0)(2\beta_0^\prime(v)-\beta_0)],
\end{equation}
with $\Omega=\sqrt{-4c_2/B_2}$, $\omega^\prime=\sqrt{4-3\beta_0/\beta_0^\prime(v)}$, $k=\sqrt{-4c_2B_2}/\hbar$,  and $\beta_0^\prime(v)$ the minimum of an effective potential, determined by the solution of
\begin{equation}
 (v+1)(v+2)=k^2\beta^{\prime3}_0(\beta^{\prime}_0-\beta_0).
\end{equation}
Inserting the specific parameters at hand, we obtain an excitation energy of 210.3~keV for the $2_1$ state ($(n_\beta,v)=(0,1)$), and the ratio $E_\textrm{ex}(4_1)/E_\textrm{ex}(2_1)=2.45$, which compares well with the numerical results (see Fig.~\ref{figure:gammaindspectrum}).  More interesting are the predictions for the first excited $v=0$ state ($(n_\beta,v)=(1,0)$).  The excitation energy of this state is predicted as 1805.1~keV, which is in reasonable agreement with the value of 1700.6~keV of the $0_3$ state depicted in Fig.~\ref{figure:gammaindspectrum}.  Here, we need to take into account that at this energy scale, we are at the limits of the approximation's validity with -4 MeV being the minimum of the original potential $V_{\gamma-\textrm{ind}}$.  Similar conclusions can be drawn for the other states in the band built upon the $0_3$ state in Fig.~\ref{figure:gammaindspectrum} and therefore, we refer to this band as the $\beta$-vibrational band of the $\gamma$-independent model, since we can approximately associate the $\beta$-vibrational quantum number $n_\beta=1$ to this band.
\subsubsection{Axially deformed rotor}
\begin{figure}[!htb]
 \includegraphics[width=\columnwidth]{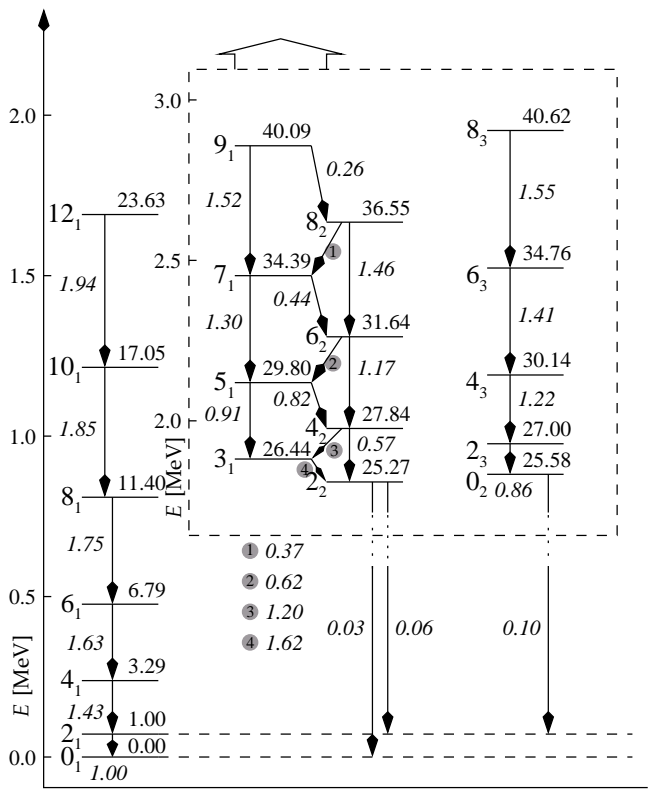}
 \caption{The energy spectrum and $B(E2)$ values of an axial rotational potential (\ref{limits:rothamiltonian}).  Energy eigenvalues are given relative to the first $L=2$ state and $B(E2)$ values relative to $B(E2;2_1\rightarrow 0_1$).}\label{figure:rotspectrum}
\end{figure}

The third limit describes axially deformed rotational structures.  In this particular case, we insert the term $[\alpha\alpha]^{(2)}\cdot\alpha$ in the Hamiltonian
\begin{equation}\label{limits:rothamiltonian}
  \hat{H}_{\textrm{rot}}=\tfrac{1}{2B_2}\pi\cdot\pi+V_\textrm{rot},
\end{equation}
breaking all the remaining degeneracies from the $\gamma$-independent rotor case.  Moreover, the classification into bands by following cascades of $B(E2)$ values is even more pronounced as the bands all occur at different energy scales in the spectrum.  Whereas the bands built on top of the $2_2$ state in the harmonic oscillator and $\gamma$-independent rotor limit had energy scales comparable with respect to the ground-state band energy scale, in this particular case, they are observed at much higher excitation energies.  This is illustrated in Fig.~\ref{figure:rotspectrum} by the dashed box, pointing out that the bands on top of the $2_2$ and the $0_2$ states are higher excited than the $12_1$ state of the ground band.

From an experimental point of view, this might be an unsatisfactory situation, as it can occur that the bands on top of the $2_2$ and $0_2$ states can appear at a much lower excitation energy and come alongside the $L=6$ or $L=8$ members of the ground band. However, the chosen potential $V_\textrm{rot}$ does not contain many free parameters to tune the excitation energy of the associated bands while keeping the minimum at a physical deformation $\beta_0\sim 0.1$.  We can as well adjust the mass parameter $B_2$ of the kinetic energy.  A decrease of the mass parameter would lead to an overall lowering of the excitation energy of the excited bands, but this would cause the ground state band to lose its rotational character.  It should be mentioned that a modified kinetic energy term was proposed in the GCM  to cope with this issue, leading towards a lowering of the excited bands in the spectrum without affecting the rotational structure of the bands.  However, we will not embark into a study of the effects of the modified kinetic energy terms in the present work, as we concentrate on the global features of the potentials, rather than the phenomenological description of experimental data.

In Fig.~\ref{figure:rotspectrum}, only the three lowest bands are depicted, with energies relative to the 71.29 keV $2_1$ state and $B(E2)$ values relative to $B(E2;2_1\rightarrow0_1)=0.01245$ in units of $(3ZeR_0^2)^2/(4\pi)^2$.
\begin{figure*}[!tb]
  \includegraphics{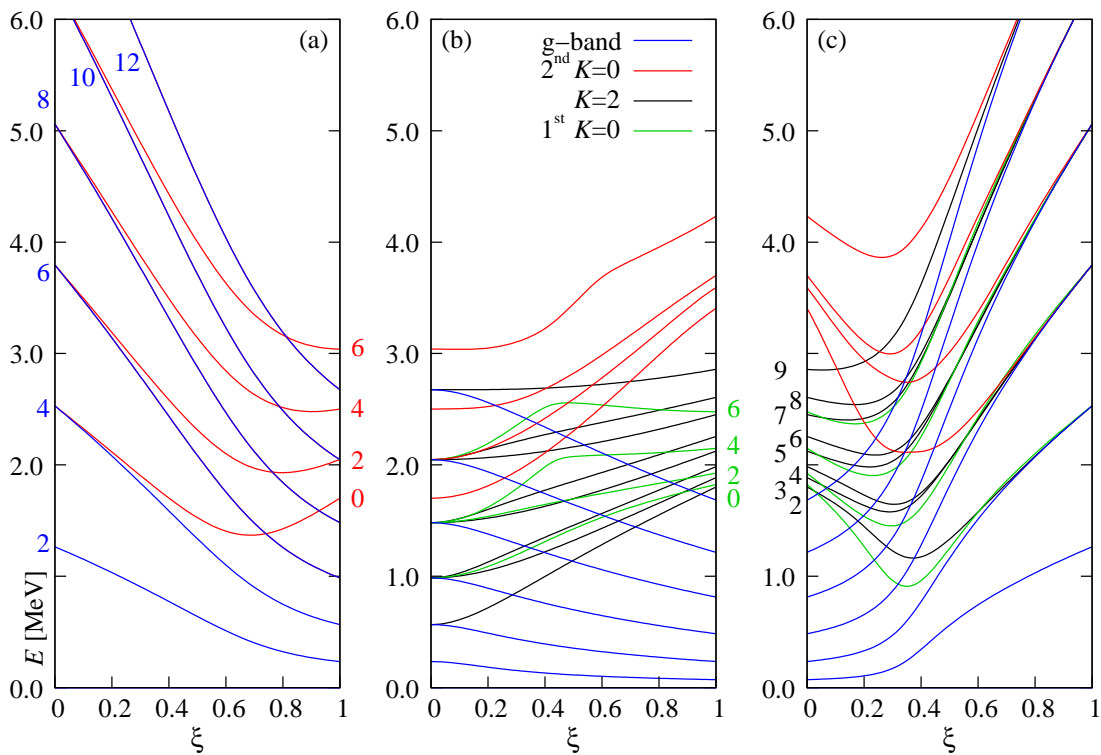}
 \caption{Energy spectra for the transitional Hamiltonians $\hat{H}_{12}$ (a), $\hat{H}_{23}$ (b) and $\hat{H}_{31}$ (c) as a function of the parameter $\xi$.}\label{figure:transspectrum}
\end{figure*}

Similar to the $\gamma$-independent rotor, the comparison of the numerical results with an analytically solvable approxmation can shed light on the general structure of the solutions.  The RVM \cite{faessler:62} is a good candidate for this purpose, as it is based on the physical assumption that an axially deformed nucleus can be described by means of a harmonic oscillator potential, in both $\gamma$ and $\beta$ around the minimum in the potential $V_\textrm{rot}$.  Consequently, we can carry out a Taylor expansion of the potential around this minimum and compare the approximate solutions with the numerical results.  These solutions can be cast in the well-known expression \cite{eisenberg:87}
\begin{align}\label{limits:rvm}
E_{LKn_2n_0}&=(\tfrac{1}{2}|K|+1+2n_2)E_\gamma+(n_0+\tfrac{1}{2})E_\beta\notag\\
&\qquad+(L(L+1)-K^2)\tfrac{1}{2}\varepsilon,
\end{align}
with $K$ the angular momentum projection quantum number along the intrinsic axis and ($n_2$,$n_0$) the vibrational quantum numbers associated with respectively the vibrations in the $\gamma$- and $\beta$-direction.  $E_\gamma$, $E_\beta$ and $\varepsilon$ are parameters, determined by the shape and localisation of the RVM potential.  Rewriting the potential $V_{\gamma\textrm{-rot}}$ around the minimum $(\beta_0,\gamma_0)=(0.264,0)$ as a Taylor expansion in $\beta$ and $\gamma$, gives rise to the following parameters of the RVM: $E_\gamma=1885.22$~keV, $E_\beta=2093.96$~keV and $\epsilon=19.15$~keV.  Substitution of these parameters in eq.~(\ref{limits:rvm}) gives some remarkable results in comparison to the solutions of $H_{\textrm{rot}}$ (see Fig.~\ref{figure:rotspectrum}).  First, the spacing of the different levels within a given band is reasonably well reproduced by the parameter $\varepsilon$.  Second, the position of the first $K=2$ band is to be expected at $E_\gamma=1888.22$~keV according to the RVM, which is in good agreement with the band built on top of the $2_2$ state in Fig.~\ref{figure:rotspectrum}.  More interesting is the classification of the first excited $K=0$ band on top of the $0_2$ state.  Within the language of the RVM model, we can either associate this band with a $\beta$-vibrational ($n_0=1,n_2=0$) or $\gamma$-vibrational ($n_0=0,n_2=1$) structure.   Substituting the corresponding quantum numbers $(n_0,n_2)$ into eq.~(\ref{limits:rvm}) gives rise to the following predictions:  the excitation energy of the $0_\gamma$ bandhead of the $\gamma$-vibrational band is to be expected at $2E_\gamma=3770.44$~keV, whereas the $0_\beta$ bandhead of the $\beta$-vibrational band can be found at $E_\beta=2093.96$~keV.  Comparison of these values with the excitation energies from the diagonalisation of $H_\textrm{rot}$ leads towards the conclusion that the lowest excited $K=0$ band is best to be associated with the $\beta$-band of the RVM.  Therefore, we will use the nomenclature of the RVM when referring to the different bands in the rotational limit $H_\textrm{rot}$.

In conclusion, the ability of the quartic truncated collective model Hamiltonian (\ref{modelspace:generalhamiltonian}) to cover vibrational, $\gamma$-independent and rotational limits has been validated through the analogy with transparent approximation schemes. This enables a clear-cut physical interpretation of the bands for each separate limit.  Also, it is clear from Figs. \ref{figure:vibspectrum} to \ref{figure:rotspectrum} that the spectra of the corresponding limits can be described respectively by the $U(5)$, the $O(6)$ and the $SU(3)$ symmetry limits of the IBM, which are the well-know vibrational, $\gamma$-independent rotor, and axial rotor limits of the IBM \cite{casten:93}. This is not too surprising as the close relation between the IBM and the BMM has long been established (see \cite{rowe:05c} and references therein), mainly based on the general characteristics of quadrupole collective models.  Intuitively, the connection can be made on the level of the potential energy (\ref{modelspace:generalpotential}) of the BMM.  This expression (\ref{modelspace:generalpotential}) can also be obtained as a total-energy surface of the IBM within the coherent state mean-field formalism \cite{vanroosmalen:82}.  From the point of view of the IBM, the ground state properties of a particular IBM Hamiltonian are encoded in the minimum of these total-energy surfaces, whereas within the BMM formalism, the minimum of the potential energy surface rather gives the mean value around which the dynamic excitations are located.  For large values of the mass parameter $B_2$ of the BMM, the ground state wavefunction is nicely confined within the potential well, with the associated energy eigenvalue approaching the minimum value of the potential as the mass parameter reaches infinity.  In the present calculations, $\hbar^2/B_2$ is chosen sufficiently small (4 keV) with respect to the typical energy scale of the potential ($\sim$ MeV).  However, we like to stress that this is an intuitive argument, rather than a hard proof for the manifestation of the connection of the respective IBM and BMM limits in Figs. \ref{figure:vibspectrum} to \ref{figure:rotspectrum}.  Presently, we will not proceed along this track as the connection is worth a study on its own and we refer the reader to recent in-depth studies \cite{garciaramos:05,garciaramos:08} on the subject.
\begin{figure*}[!htbp]
	\includegraphics{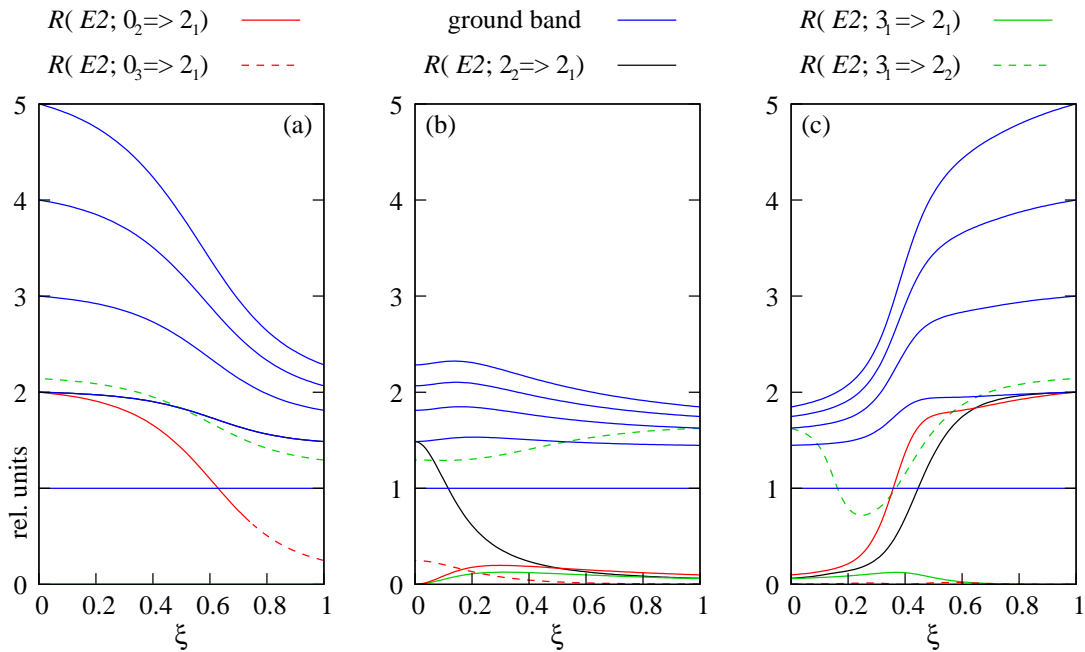}
	\caption{Selected $B(E2)$ values for the transitional Hamiltonians $\hat{H}_{12}$ (a), $\hat{H}_{23}$ (b) and $\hat{H}_{31}$ (c) along the transition lines, relative to the $B(E2;2_1\rightarrow0_1$) value (see equation (\ref{transition:transrelbe2})). }\label{figure:transrelbe2}
\end{figure*}

\subsection{Transition paths}
\begin{figure*}[!htb]
	\includegraphics{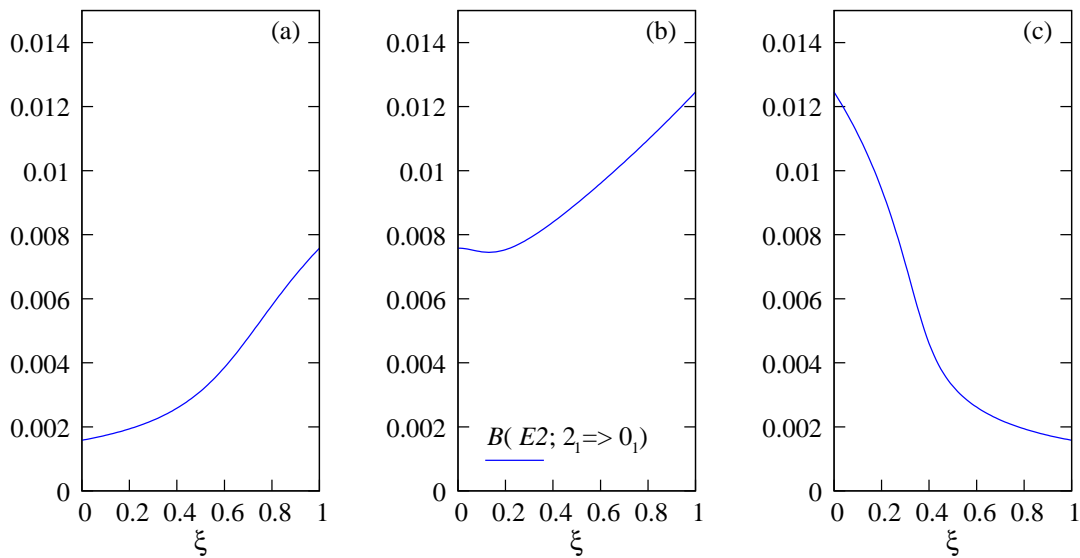}
	\caption{Absolute $B(E2;2_1\rightarrow0_1)$ values for the transitional Hamiltonians $\hat{H}_{12}$ (a), $\hat{H}_{23}$ (b) and $\hat{H}_{31}$ (c) along the different transition paths, given in units $(3ZeR_0^2)^2/(4\pi)^2$.}\label{figure:transabsbe2}
\end{figure*}

\begin{figure}[!htb]
	\includegraphics{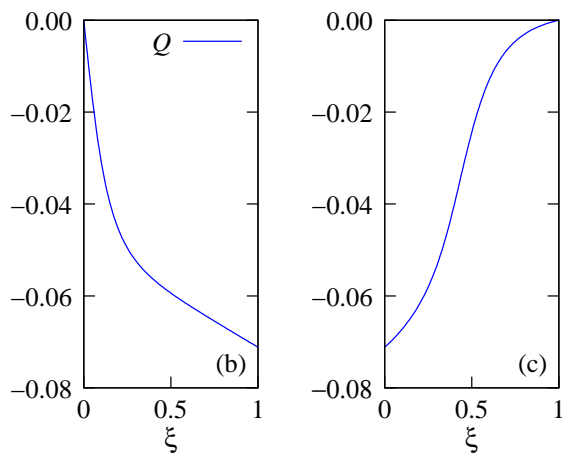}
	\caption{The quadrupole moments $Q$ of the first excited $2_1$ state, given in units $\sqrt{16\pi/5}(3ZeR_0^2)/(4\pi)$.  Only the values for the transitional Hamiltonians $\hat{H}_{23}$ (b) and $\hat{H}_{31}$ (c) are presented as the quadrupole moments for the harmonic oscillator to the $\gamma$-independent rotor are identically zero.}\label{figure:transquadmom}
\end{figure}

Having determined the structure of the typical limiting cases in the collective model, it is interesting to see how these structures evolve as the potentials gradually change from one limiting case into another.  Therefore, we construct the following transitional Hamiltonians
\begin{align}
 \hat{H}_{12}&=\tfrac{\hbar^2}{2B_2}\pi\cdot\pi+(1-\xi)V_\textrm{spher}+\xi V_{\gamma-\textrm{ind}},\label{transition:transhamH12}\\
\hat{H}_{23}&=\tfrac{\hbar^2}{2B_2}\pi\cdot\pi+(1-\xi)V_{\gamma-\textrm{ind}}+\xi V_\textrm{rot},\label{transition:transhamH23}\\
\hat{H}_{31}&=\tfrac{\hbar^2}{2B_2}\pi\cdot\pi+(1-\xi)V_\textrm{rot}+\xi V_\textrm{spher},\label{transition:transhamH31}
\end{align}
for which we calculate the excitation energy and electric quadrupole observables along the transition path.  The results are presented in Figs. \ref{figure:transspectrum}, \ref{figure:transrelbe2}, \ref{figure:transabsbe2} and \ref{figure:transquadmom}.  For easy comparison, the three different transition paths (respectively $\hat{H}_{12}$, $\hat{H}_{23}$ and $\hat{H}_{31}$) are plotted side by side with the same scaling.  It should be mentioned that the study of $\hat{H}_{12}$ is closely related to recent work on phase transitions in the Bohr-Mottelson model \cite{turner:05a}, where the Hamiltonian is pushed into the domain of large deformations in order to clearly identify the critical point in the transition and discuss the apparent quasi-dynamical symmetry along the transition line.

We start with the discussion of the energy spectra.   In Fig.~\ref{figure:transspectrum}, the energy spectra are plotted as a function of $\xi$.  In the left panel (a), the transition from the spherical to the $\gamma$-independent rotor is presented (see eq.~\ref{transition:transhamH12}).  Since $SO(5)$ is a symmetry for both limits, it is also a symmetry for every intermediate Hamiltonian $\hat{H}_{12}$, which makes the seniority quantum number $v$ a good quantum number along the whole path.  Furthermore, the large degeneracy of the harmonic oscillator persists along the transition path, except for the band on top of the $L=0$ state originating from the $0_2$ state of the harmonic oscillator (see Fig.~\ref{figure:vibspectrum}).  One can clearly see from Fig.~\ref{figure:transspectrum}(a) how this band (depicted in red) decouples from the ground band (in blue) to become the $\beta$-vibrational band in the $\gamma$-independent limit.   The different nature of this band is even more apparent from the pure crossings around $\xi\sim0.8$.  Similarly, one can deduce from the relative $B(E2)$ values related to this band (see Fig.~\ref{figure:transrelbe2}(a)) that this $\beta$-vibrational like $L=0$ ($0_2$ up to $\xi\sim 0.8$) state differs structurally from the other one ($0_3$ up to $\xi\sim 0.8$), as the $B(E2;0_2\rightarrow2_1)/B(E2;2_1\rightarrow0_1)$  value decreases faster than the values $B(E2;L_1\rightarrow(L-2)_1)/B(E2;2_1\rightarrow0_1)$ of the ground-band members or even the $B(E2;3_1\rightarrow2_2)/B(E2;2_1\rightarrow0_1)$.  To simplify the notation, it is convenient to introduce the following quantity
\begin{equation}\label{transition:transrelbe2}
R(E2;L_i\rightarrow L_f)=\frac{B(E2;L_i\rightarrow L_f)}{B(E2;2_1\rightarrow0_1)},
\end{equation}
for the $B(E2)$ values relative to $B(E2;2_1\rightarrow0_1)$ as plotted in Fig.~\ref{figure:transrelbe2}.

In the middle panel (b) of Fig.~\ref{figure:transspectrum}, the transition from the $\gamma$-independent to axially deformed rotor is plotted (see eq.~\ref{transition:transhamH23}).  By introducing the seniority breaking term $[\alpha\alpha]\cdot\alpha$ in the Hamiltonian, the remaining $SO(5)$ degeneracy of the $\gamma$-independent rotor is lifted.  This effect arises rather promptly with small values of $\xi$, leading to the instant development of separate bands in the spectrum.  At the right-end of the transition path ($\xi=1$), a clear picture takes on form with distinct rotational and vibrational-like bands at different energy scales.  The fact that the states can now unambiguously be organised into bands is also reflected in the $R(E2;L_i\rightarrow L_f)$ values in the middle panel (b) of Fig.~\ref{figure:transrelbe2}.  There it can be seen that the value $R(E2;2_2\rightarrow 2_1)$ quickly drops whereas $R(E2;3_1\rightarrow 2_2)$ stays reasonably unaffected along the transition line, which clearly points out that $2_2$ and $3_1$ belong to the same ($K=2$) band, depicted in the dashed box of Fig.~\ref{figure:rotspectrum}.

In the right panel (c) of Fig.~\ref{figure:transspectrum}, we close the circle with the transition from the axially deformed limit to the vibrational limit.  It can be seen from that figure that this transition is less gradual, compared to the harmonic oscillator to $\gamma$-independent rotor transition.  In the latter case, only the $SU(1,1)$ symmetry was broken, whereas in the former case also the $SO(5)$ symmetry is immediately broken by adding the $[\alpha\alpha]^{(2)}\cdot\alpha$ term in the Hamiltonian.

Before proceeding, there is a peculiarity to be noted concerning the identification of the $K=0$ bands in Fig.~\ref{figure:transspectrum} as either a $\gamma$- or $\beta$-vibrational band.  From the comparison of the axially deformed limit with the predictions made by RVM, we could unambiguously identify the lowest $K=0$ (built on the $0_2$ state) as the $\beta$-vibrational band.  This can further be justified if we follow the transition path towards the harmonic oscillator, as depicted in the right panel (c) of Fig.~\ref{figure:transspectrum}.  We notice that, {\it e.g.}~the $0_2$ state of this $\beta$-vibrational band (green lines) evolves towards the $(n=1,v=0)$ $SU(1,1)$ representation of the harmonic oscillator, which is basically a $\beta$-vibration.  However, this is not the only path to reach the harmonic oscillator.  One could go the opposite way in Fig.~\ref{figure:transspectrum}, and follow the transition from axially deformed rotor to harmonic oscillator limit via the $\gamma$-independent rotor limit.  Following {\it e.g.}~the $0_2$ state along each transition (middle (b) and left panel (a)), we find that in this case, we end up in the $(n=0,v=3)$ state, which cannot be associated with $\beta$-vibrations.  The solution to this paradox lies in the middle panel (b).  From this panel, it is clear that all states in the $\gamma$-independent to axial rotor limit are subject to considerable mixing and 'no-crossing' effects, due to the large admixture of states with different seniority quantum numbers.  Moreover, as the $0_3$ state of the $\gamma$-independent rotor limit clearly evolves from the $0_2$ $(n=1,v=0)$ state of the harmonic oscillator, we can state that the $0_2$ and $0_3$ states must have switched nature along the transition path from $\gamma$-independent to axially deformed rotor.  In conclusion, the association of the $K=0$ bands with $\beta$- or $\gamma$-vibrational structures should be treated with caution as considerable mixing effects can arise, perturbing the simple picture of vibrational motion along the $\beta$- or $\gamma$ direction.  

Apart from the relative $B(E2)$ values, it is also interesting to calculate the absolute $B(E2)$ values, as they give an indication of the collective deformation of an atomic nucleus \cite{eisenberg:87}.  
In Fig.~\ref{figure:transabsbe2}, the $B(E2;2_1\rightarrow0_1)$ values are presented along the transition paths.  One can clearly see that, going from the harmonic oscillator to the $\gamma$-rotational limit, the $B(E2;2_1\rightarrow0_1)$ value rises steadily, due to the manifestation of a deformed minimum in the potential.  This rise persists in the transition from the $\gamma$-independent rotor to the rotational limit, as the onset of the term $[\alpha\alpha]\cdot\alpha$ in the potential breaks the symmetry in the $\gamma$-direction, driving the minimum in the potential towards prolate structures.  Finally, along the rotational to vibrational transition path, the $B(E2;2_1\rightarrow0_1)$ drops back to the originally value of the spherical harmonic oscillator.

Similar conclusions can be drawn from a study of the spectroscopic quadrupole moment of the first excited $2_1$ state, defined by 
\begin{align}
Q=\sqrt{\tfrac{16\pi}{5}}\langle 2_1 M=2|\hat{T}(E2)|2_1M=2\rangle,
\end{align}
where the same linear approximation of $\hat{T}(E2)$ is used as for the calculation of the $B(E2)$ values (see eqs.~(\ref{limits:be2}) and (\ref{limits:te2})).  The results for $Q$ are presented in Fig.~\ref{figure:transquadmom}.  Contrary to the other figures, the transition from the harmonic oscillator to the $\gamma$-independent rotor is not depicted since, within the linear approximation of $\hat{T}(E2)$, the selection rules for $\alpha$ ($\Delta v=\pm1$) render the quadrupole moments identically zero along this whole transition path.  Here again, we notice that the quadrupole moment sharply rises (in absolute value), as soon as the seniority-breaking term in the potential is turned on (along both transition paths), to reach a maximum at the rotational limit.  This rise is more pronounced in the transition from the $\gamma$-independent rotor to rotational limit (b) as compared to the transition from the spherical to rotational limit (c).  This difference stems from the fact that in the former transition the potential already exhibits a deformed minimum in the $\beta$ direction, while in the latter transition the minimum in the $\beta$ direction is steadily developing from the spherical minimum towards a minimum at a distinct deformation $\beta_0$.
\section{Conclusions and outlook}\label{section:conclusions}

In the present paper, we have studied the spectral properties of a truncated collective Hamiltonian up to quartic terms in the potential energy.  This Hamiltonian is able to cover three different limits which can be associated with vibrational, $\gamma$-independent rotational and rotational structures.  This has been demonstrated by comparison of these limits with physically transparent approximation schemes and the connection with the three branching limits of the IBM.  Furthermore, it has been studied how the typical fingerprints of collective structure (such as \textit{e.g.} energy spectra, $B(E2)$-values, etc.) evolve if the Hamiltonian changes from one particular limit to another, which gave rise to some remarkable effects due to seniority mixing in the $\gamma$-independent rotational to axially deformed rotational limit.

The present results also provide a genuine test for the recently developed program which treats the (general) collective model within a Cartan-Weyl framework.  This is necessary if one wants to proceed towards more complex collective structures, such as \textit{e.g.} triaxiality and shape coexistence, which are prominently present in the medium-heavy to heavy isotopes of the nuclear chart.  Extending the present truncated Hamiltonian to incorporate higher-order terms will be the subject of future investigation.

\begin{acknowledgments}
 The authors would like to acknowledge financial support from the ''FWO Vlaanderen'', Ghent University and the Interuniversity Attraction Pole (IUAP) under projects P5/07 and P6/23.  S.D.B. acknowledges a travel grant from the ''FWO Vlaanderen'' supporting a stay at the University of Toronto.
\end{acknowledgments}
\bibliography{debaerdemacker_collectivehamiltonian}
\bibliographystyle{h-physrev}
\newpage
\end{document}